# 0-LEVEL VACUUM PACKAGING RT PROCESS FOR MEMS RESONATORS

*Nicolas Abelé[1,3], Daniel Grogg[1], Cyrille Hibert[2], Fabrice Casset[4], Pascal Ancey[3], Adrian* M. Ionescu[1]

[1]LEG, Ecole Polytechnique Fédérale de Lausanne (EPFL), Switzerland, [2]CMI (EPFL), [3]ST Microelectronics, France, [4]CEA-LETI MINATEC, France

**ABSTRACT**

A new Room Temperature (RT) 0-level vacuum package is demonstrated in this work, using amorphous silicon (aSi) as sacrificial layer and $SiO_2$ as structural layer. The process is compatible with most of MEMS resonators and Resonant Suspended-Gate MOSFET [1] fabrication processes. This paper presents a study on the influence of releasing hole dimensions on the releasing time and hole clogging. It discusses mass production compatibility in terms of packaging stress during back-end plastic injection process. The packaging is done at room temperature making it fully compatible with IC-processed wafers and avoiding any subsequent degradation of the active devices.

## 1. INTRODUCTION

MEMS resonators performances have been demonstrated to satisfy requirements for CMOS co-integrated reference oscillator applications [2-3]. Different packaging possibilities were proposed in previous years using either a 0-level approaches [4, 5] or wafer bonding approaches [6]. According to industry requirements, 0-level thin film packaging using standard front-end manufacturing processes is however likely to be the most cost-efficient technique to achieve vacuum encapsulation of MEMS components for volume production.

## 2. DEVICE DESCRIPTION AND PACKAGING DESIGN

The packaging process has been done on a MEMS resonator having MOSFET detection [1]. The device is based on a suspended-gate resonating over a MOSFET channel which modulates the drain current. The advantage of this technique is the much larger the output detection current than for the usual capacitive detection type, due to the intrinsic gain of the transistor.

The RSG-MOSFET device fabrication process and performances were previously described in [7]. The process steps are presented in Fig. 1, where a 5µm thick amorphous silicon (aSi) layer is sputtered on the already released MEMS resonator followed by a 2µm RF sputtered $SiO_2$ film deposition. A quasi-zero stress aSi film deposition process has been developed; the quasi-vertical deposition avoids depositing material under the beam lowering the releasing time. Releasing holes of 1.5µm were etched through the $SiO_2$ layer and the releasing step is done by dry $SF_6$ plasma. Due to pure chemical etching, high selectivity of less than 1nm/min on $SiO_2$ was obtained. The holes were clogged by a non-conformal sputters $SiO_2$ deposition at room temperature.

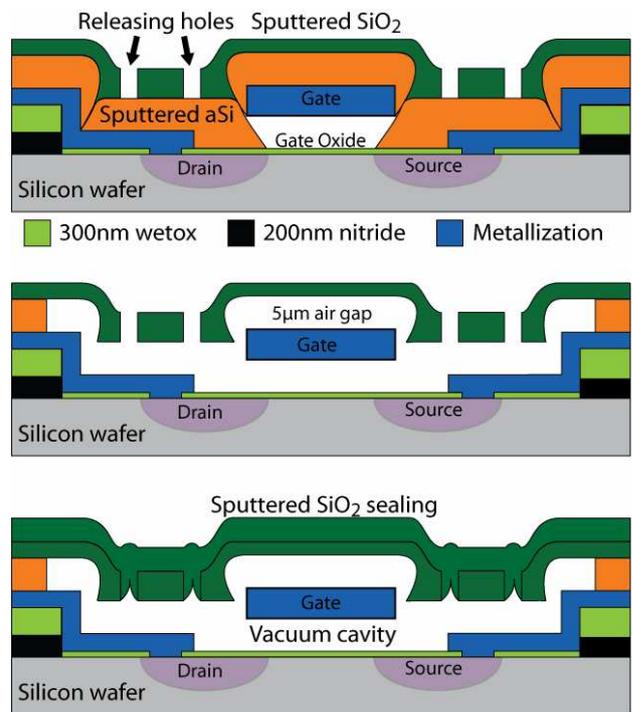

*Fig. 1 Schematic of the 0-level vacuum package fabrication process of a RSG-MOSFET*

Packaging process has been performed on the metal-gate SG-MOSFET and Fig. 2a shows an SEM picture of a released AlSi-based RSG-MOSFET with a 500nm air-gap, a beam length and width of respectively 12.5µm and 6µm with a 40nm gate oxide. A vacuum packaged RSG-MOSFET is shown in Fig. 2b highlighting the strong bonds of the re-filled releasing hole after clogging. Cross section of a releasing hole in Fig. 2c shows more than 1µm bonding surface to ensure cavity sealing. A FIB cross section in Fig. 2d shows the suspended $SiO_2$ membrane above the suspended-gate. The vacuum atmosphere inside the cavity is obtained by depositing the top $SiO_2$ layer under $5 \times 10^{-7}$ mBar given by the equipment.

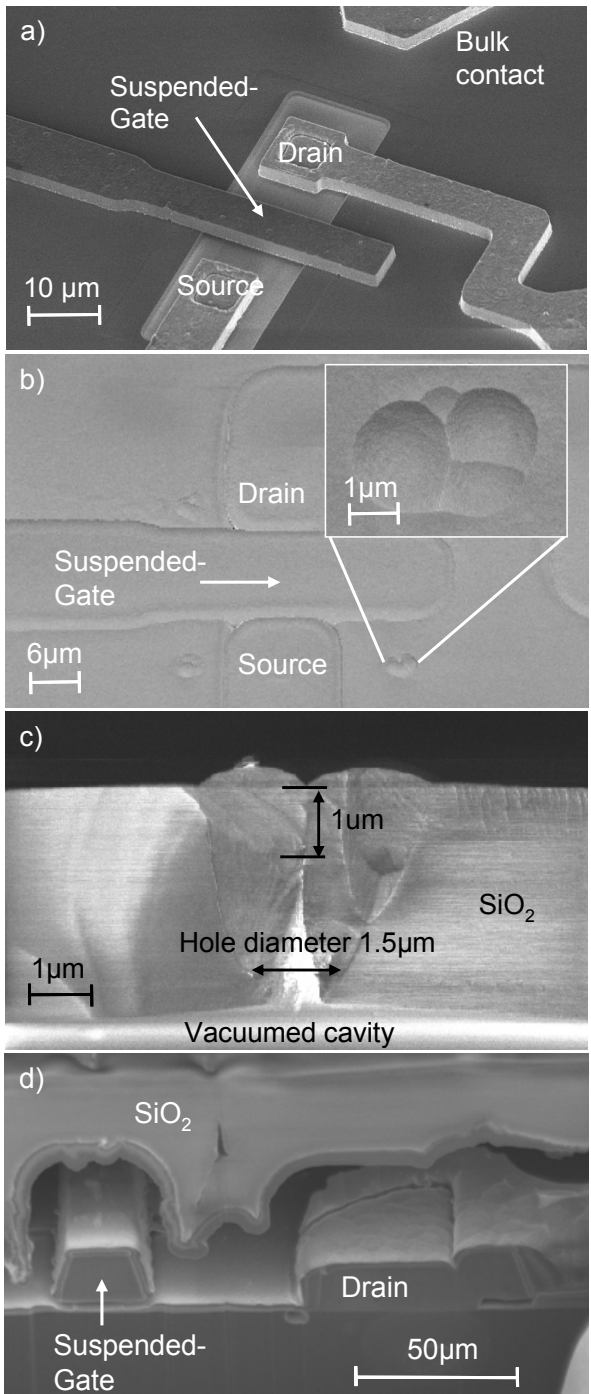

*Fig. 2 SEM pictures of a) AlSi-based RSG-MOSFET, b) Top view of a $SiO_2$ cap covering the RSG-MOSFET, c) Cross section of releasing holes filled with sputtered $SiO_2$, d) FIB cross section of the packaged RSG-MOSFET, material re-deposited during the FIB cut is surrounding the suspended-gate and the $SiO_2$ membrane.*

The slightly compressive $SiO_2$ membranes show very good behavior for the thin film packaging, as seen in Fig.3 where cavities were formed on large opening size. During the clogging process, due to the highly non-conformal deposition, the amount of material entering in the cavity has been measured to be only 80nm compared to the 2.5μm oxide deposited. Residues inside the cavity are confined in an 8-to-10μm diameter circle, but strongly depend on the topology inside the cavity. The oxide thickness needed to clog the holes strongly depends on the hole width-over-height ratio, which therefore determines the amount of residues in the cavity.

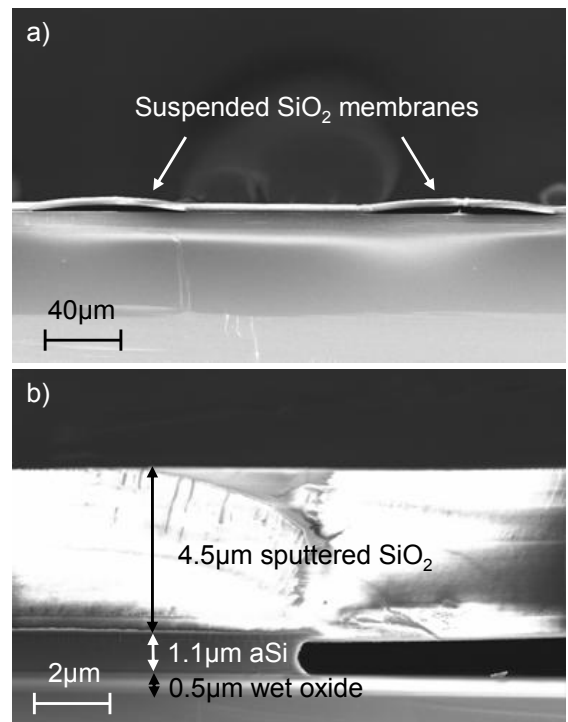

*Fig. 3 a-b) Cross section of a 2um $SiO_2$ suspended membrane having a releasing hole clogged by a 2.5μm $SiO_2$ sputtering deposition*

## 3. EFFECT OF OPENING SIZE ON RELEASING RATE AND CLOGGING EFFECT

Etching rate variation on aSi related to the hole opening size and the aSi thickness is shown in Fig. 4. Small holes openings decrease the etching rate. A dual underetching behavior due to aSi thickness variation and holes diameters is observed after a 2 min. release step: for a small hole aperture (2μm diameter), exposed surface factor is dominant and etching rate is 3 times greater for the thin aSi. However for large openings (9μm diameter) for which underetch distance is more important, path factor representing the lateral opening height for species to reach aSi becomes important and then etching ratio decreases to 1.3.

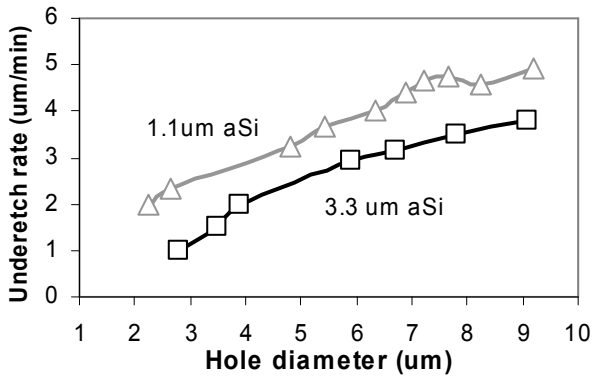

*Fig. 4 Underetch rate for various releasing holes diameters with amorphous silicon sacrificial layers of 1.1μm and 3.3μm, after 2min. releasing.*

After release, encapsulation is performed by sputtered deposition of $SiO_2$ under high vacuum of $5 \times 10^{-7}$ mbar using the intrinsic, non-conformal deposition to clog holes, as shown in Fig. 5. Clogging effect is strongly material dependent and is related to the sticking coefficient that defines probability for a molecule to stick to the surface. The coefficient is below 0.01 for LPCVD Poly-Si but 0.26 for $SiO_2$, therefore being more suitable for clogging purpose.

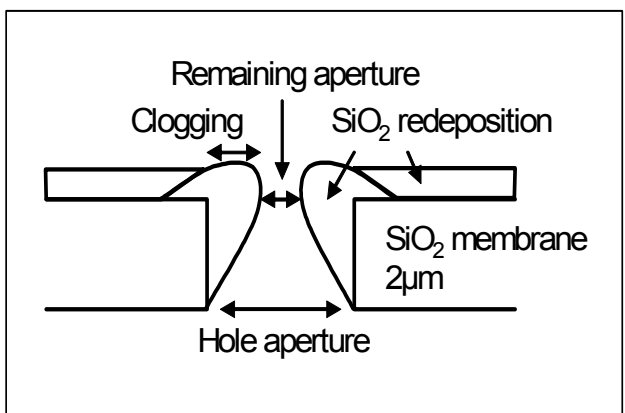

*Fig. 5 Schematic of a cross section of the $SiO_2$ membrane clogged by $SiO_2$ sputtering deposition*

Hole clogging has a strong dependence on the opening aspect ratio as presented in Fig. 6. Holes with diameter-over-height aspect ratio below 1 are clogged for $SiO_2$ thickness of 2μm. Hole with opening ratio of 1.5 could only be clogged for a 3μm thick $SiO_2$ deposition. The hole clogging rate is measured to be 330nm per deposited micron of $SiO_2$.

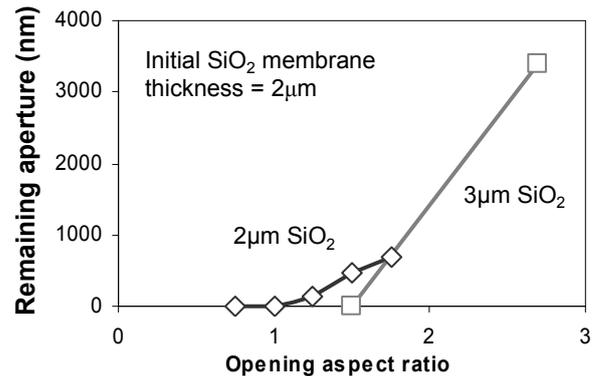

*Fig. 6 hole clogging effect depending on the diameter-over-height ratio in the 2μm $SiO_2$ membrane (Right). Remaining aperture diameter (in nm) for 2μm and 3μm $SiO_2$ deposition for hole clogging.*

The effect of hole geometry on underetch rate and clogging has been studied on square and rectangular holes in Fig.7. Rectangular opening has a quasi identical underetching than square shape of the same opening area, while clogging is 10 times more important.

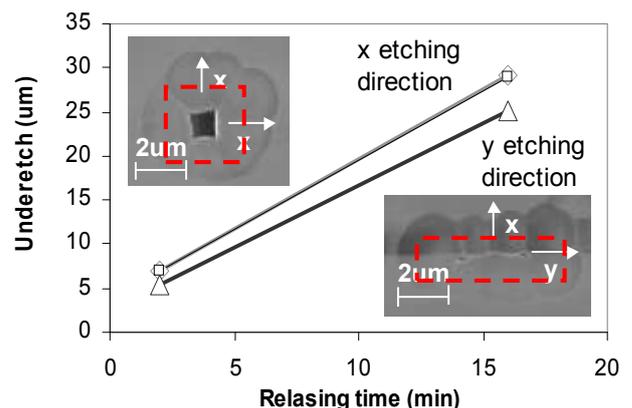

*Fig. 7 Underetch length after 16min release for 29.1μm2 square and rectangle release holes (red dotted rectangles).*

The initial $SiO_2$ thickness is a 2μm and the thickness of aSi is 1.1μm. Remaining hole size after 2.5μm $SiO_2$ deposition is 1.4μm for the square and 140nm for the rectangle.

### 4. PACKAGING ISSUES FOR PRODUCTION ENVIRONMENT

For industrial production of integrated MEMS, 0-level package has to sustain plastic molding, which corresponds to an isostatic pressure of around 100Bar. Encapsulation film thickness has been designed to lower the impact of the pressure during molding. FEM simulations done with Coventor® in Fig. 8 show that the molding-induced package deflection is reduced to 25nm, having a 4.5μm thick $SiO_2$ film, which makes it compatible with standard industrial back-end processes.

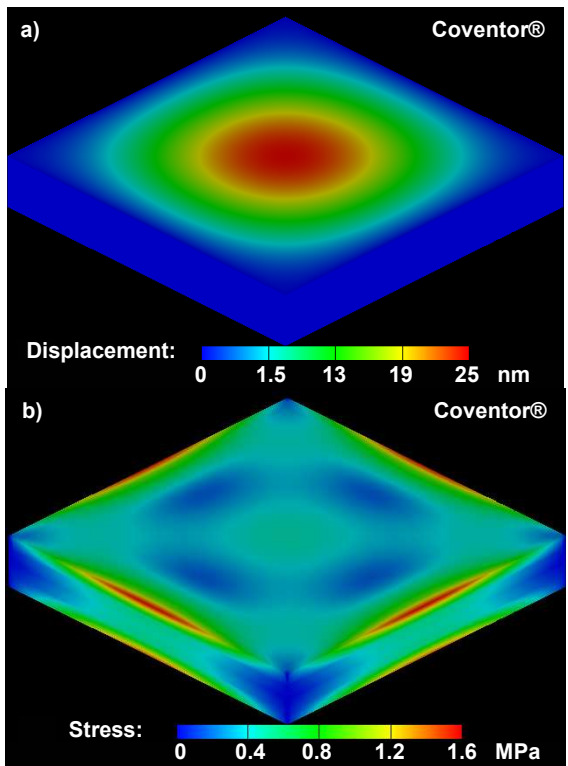

*Fig. 8 FEM modelling of the packaged resonator under applied isostatic pressure mimicking plastic injection process step.*

Effect of LTO and PECVD nitride materials on capping deflection under molding stress are presented in Table I. Membrane thickness can then be optimized to lower the molding-induced deflection by considering Young's modulus and maximum stress before failure of the two materials.

| Structural layer material | LTO | Nitride PECVD |
|---|---|---|
| Film thickness | 4.5μm | 2.5μm |
| Max. stress before failure | 2GPa | 9GPa |
| Stress due to molding | 1.6MPa | 4MPa |
| Molding-induced deflection | 25nm | 36nm |

*Table I. FEM simulations of the structural layer thickness needed to sustain plastic molding over 0-level packaging composed of a 30μmx30μm membrane. Comparison with PECVD nitride thickness needed to induce the same deflection.*

On the developed process flow, further investigations on vacuum level and long term stability still to be studied in order to fully characterize the packaging. This characterization can either be done directly by using helium leakage test [9], or indirectly by actuating the packaged resonator for which quality factor is directly related to the vacuum level.

## 5. CONCLUSION

A novel 0-level packaging process was presented using aSi as sacrificial layer and $SiO_2$ as encapsulating layer. RSG-MOSFET resonators have been successfully encapsulated under high vacuum. Impact of back-end-of-line industrial process over the encapsulation has been investigated, resulting in optimal cover thickness needed to sustain plastic molding. Influence of hole dimensions on releasing time and clogging effect for encapsulation were investigated, and optimized packaging parameters are identified for this process.

## 11. REFERENCES

[1] N. Abelé et al., "Ultra-low voltage MEMS resonator based on RSG-MOSFET ", MEMS '06, pp. 882-885, 2006

[2] V. Kaajakari et al., "Low noise silicon micromechanical bulk acoustic wave oscillator", IEEE International Ultrasonics Symposium, pp. 1299- 1302, 2005

[3] Y.-W. Lin et al., "Low phase noise array-composite micromechanical wine-glass disk oscillator," IEDM '05, pp. 287-290, 2005

[4] N. Sillon et al., Wafer Level Hermetic Packaging for Above-IC RF MEMS: Process and Characterization, IMAPS 2004

[5] B. Kim et al.,, "Frequency Stability of Wafer-Scale Encapsulated MEMS Resonators," Transducers '05, vol. 2, pp. 1965-1968, 2005

[6] V. Kaajakari et al., "Stability of wafer level vacuum encapsulated single-crystal silicon resonators", Sensors and Actuators A: Physical, Vol. 130-131, pp. 42-47, 2006

[7] N. Abelé et al., "Suspended-Gate MOSFET: bringing new MEMS functionality into solid-state MOS transistor ", IEDM '05, LATE NEWS, pp. 479-481, 2005

[8] S. Frédérico et al.,"Silicon sacrificial layer dry etching (SSLDE) for free-standing RF MEMS architectures", MEMS '03, pp. 570- 573, 2003

[9] I. D. Wolf at al., "The Influence of the Package Environment on the Functioning and Reliability of Capacitive RF-MEMS Switches," Microwave Journal, vol. 48, pp. 102-116, 2005.